\documentstyle[dina4,fleqn,espcrc2]{article}
\input psfig

\title{ Pair fluctuation effects above $T_{\rm c}$}
\author{{\bf J.J. Rodr\'{\i}guez-N\'u\~nez}$^{1,4}$,
S. Schafroth$^{2,4}$, R. Micnas$^3$, T. Schneider$^4$,
H. Beck$^1$ and M.H. Pedersen$^{2,4}$\\
$^1$Institut de Physique,
Universit\'e de
Neuch\^atel,
Rue A.L. Breguet 1, CH-2000 Neuch\^atel,
Switzerland\\
$^2$Physik-Institut der Universit\"at
Z\"urich, Winterthurerstr.\ 190,
CH-8057 Zurich, Switzerland \\
$^3$Institute of Physics, Adam Mickiewicz University,
Solid State Division, Matejki 48/49, PL-60-769 Pozna\'n, Poland \\
$^4$IBM Research Division, Zurich Research Laboratory,
CH-8803 R\"uschlikon, Switzerland}

\begin{document}

\begin{abstract}
We explore the occurrence of pairing effects above
$T_{\rm c}$ in the 2D
attractive Hubbard model. The presence
of pairs above $T_{\rm c}$ goes
beyond the BCS approximation, in which
pair formation and condensation
occur at the same temperature. Using the fully
self-conserving $T$-matrix formalism, which is valid
above $T_{\rm c}$, we find that (1) the distribution function,
$n(k)$, shows a 10\% change
of the weight from ``below'' to ``above''
$k_{\rm F}$ with respect to the free case, and (2) the phase shift,
$\delta\phi(\omega,k)$, shows
$\Theta$-like behavior as a function of $\omega$
for large momentum $k$ along the diagonal
of the Brillouin zone. Our
calculations have been carried out for an
interaction of $U/t = - 4.0$ and a
temperature of $T/t = 0.125$, where $t$
is the hopping matrix element between nearest
neighbors. We conclude
that for such an interaction value
the Fermi surface is not a well-defined quantity.
\end{abstract}

\maketitle
The high-$T_{\rm c}$ materials with their short coherence
length and their
large penetration depth are leading condensed-matter physicists to
consider the effect of pair fluctuations above $T_{\rm c}$, i.e., to
investigate the exotic properties of the normal
state. To include
fluctuations above $T_{\rm c}$, we must go beyond
the BCS approach, which
is mean-field approximation, and where, therefore,
pair formation and
superconductivity occur at the same temperature. The next
level of approximation to BCS is the fully
self-consistent $T$-matrix formalism, which includes
the effect of fluctuations
in a natural way by means of the self-energy of the system.
The effect of fluctuations invalidates mean-field theory
close to $T_{\rm c}$; but away from $T_{\rm c}$, the pronounced
uniaxial
anisotropy of cuprates leads to a crossover
from three dimensions
(3D) to quasi-2D behavior, where fluctuations
are again essential
\cite{S1}. We explore the pair fluctuation effects above
$T_{\rm c}$ in the 2D attractive Hubbard
model \cite{MRR} within
the fully self-consistent $T$-matrix formalism \cite{KB}.
Our basic assumption
is that there are fluctuation pairs above
the critical temperature and that at $T_{\rm c}$ there is
Bose-Einstein condensation of pairs. The signaling of
an order parameter in the $T$-matrix approach is given
by the divergence of the real part of the $T$-matrix
evaluated at zero momentum and zero frequency (Thouless
criterion).

We have evaluated the distribution function, $n({\bf k})$,
for an interaction of $U/t = -4.0$
along the diagonal of the Brillouin zone, and compare it
with the distribution function for $U/t = 0.0$.  The
temperature chosen is $T/t = 0.125$, and the total
density is $n = 0.32$. These two
distribution functions are shown in Fig. 1.
We have gone up to 11 points of the Brillouin
zone, with the wave number in units of ${\pi Q}/{16}$. The parameter
$\alpha$, as defined in Ref.\ \cite{PN},
measures the relative
number of excited quasiparticles. Here
$\alpha$ is equal to $20\%$, which indicates a considerable
renormalization of the distribution function with respect to
the non-interacting case. Also, we observe that the
distribution function, instead of being sharp, is rather
smeared out. This leads us to conclude that the Fermi
surface is not a well-defined quantity when correlations
are relevant. This distribution function
can be fitted with a fluctuating BCS gap, i.e.,
by an auxiliary fluctuating field, $\Delta$,
in the partition function.

Next we calculated the phase shift, $\delta
\phi (\omega)$, for large values of momenta along the
diagonal of the Brillouin zone. The results are
presented in Fig. 2 for a density
of $n = 0.20$. We observe that for higher
momenta the phase shift tends to
saturate \cite{NSR,Varma}. The appearance of
$\Theta$-like behavior for the phase shift is an indication of
resonant or bound states.
Their contribution can be estimated in
the ring approximation for the thermodynamic
potential, which gives one density contribution
that stems purely from band effects and one that is
due to bound or resonant states. We have
worked with 32$\times$32 points in the
Brillouin zone and 1024 Matsubara frequencies.

In conclusion, we have shown that, when correlations are important,
the inclusion of
pair fluctuations above $T_{\rm c}$ in
the fully self-consistent $T$-matrix formalism
produces considerable
renormalization of the distribution function. The fact
that the distribution function is not as sharp as
in the free case leads us to conclude that the Fermi
surface is not a well-defined quantity for a
system in which correlations are important. At
the same time, we obtain the signatures of resonant or
bound states by studying the phase shifts for large
momenta and large energies. We should add that the
$T$-matrix formalism, for the case of diagonal
Green's function, should be used for temperatures up to
the critical temperature, $T_{\rm c}$, which is given
by the divergency of the real part of the $T$-matrix at
zero frequency and zero
momentum.

\vskip 1mm

We gratefully acknowledge the support of the Swiss
National Science Foundation. RM acknowledges partial
support from the Committee for Scientific Research (KBN
Poland, project No. 2 P3 02 057 04). JJRN thanks for partial
support from the project N$^o$. F-139 (CONICIT) and
CONDES-LUZ. We would also like to thank Mar\'{\i}a D. Garc\'{\i}a
for reading the manuscript.

\begin{figure}
\vspace*{4mm}
\caption{ \label{fig1} The momentum distribution
function along the diagonal of the Brillouin zone.}
\end{figure}
\begin{figure}
\caption{ \label{fig2}
The phase shift as function of frequency for large momenta,
$Q$, along the diagonal of the Brillouin zone ($32 \times 32)$.
The $Q$-values are integers.}
\end{figure}

\end{document}